\documentclass[reqno]{amsart}
\usepackage{amsmath,amsthm,latexsym,amssymb,amscd}
\usepackage{amsfonts}
\usepackage{graphicx,graphics}
\usepackage[french,spanish,english]{babel}
\usepackage[latin1]{inputenc}
\font\tit=cmbx10 scaled \magstep3 
\font\aut=cmcsc10 scaled \magstep1 \font\afilia=cmr10 scaled
\font\smcc=cmcsc8

\numberwithin{equation}{section}

\begin{document}

\title[Solving the seventh-order Kaup-Kupershmidt equation]{}

\author[Alvaro Salas, Jairo Castillo and Gonzalo Escobar]{}\label{pagini}

\date{}
\maketitle \setcounter{page}{1}

\centerline{\tit About the seventh-order Kaup-Kupershmidt equation}

\smallskip
\centerline{\tit  and its solutions}

\vspace{0.8cm} \centerline{\aut Alvaro H. Salas S.  }
\centerline{ \emph{email} : asalash2002@yahoo.com}
\smallskip
\centerline{ Department of Mathematics} \centerline{\afilia
Universidad de Caldas, Manizales, Colombia.}
 \centerline{\afilia Universidad
Nacional de Colombia, Manizales.}

\vspace{0.3cm} \centerline{\aut Jairo Ernesto Castillo Hern\'andez}
\smallskip
\centerline{ Department of Physics} \centerline{\afilia
Universidad Aut\'onoma, Bogot\'a, Colombia.}
\centerline{\afilia
Universidad Central, Bogot\'a, Colombia.}

\vspace{0.3cm} \centerline{\aut Jos\'e Gonzalo Escobar Lugo}
\centerline{ \emph{email} : jogoel@gmail.com}
\smallskip
\centerline{\afilia
Universidad Cooperativa de Colombia, Bogot\'a.}
\centerline{ Department of Mathematics} \centerline{\afilia
Universidad Libre, Bogot\'a, Colombia.}

\bigskip
{\narrower \noindent {\smcc Abstract.} \small In this letter we
obtain exact soliton and periodic solutions to the seventh-order
Kaup-Kupershmidt equation. We make use of the Cole-Hopf
transformation and two particular rational hyperbolic functions
ansatze.
\par}

\smallskip
{\narrower \noindent {\itp Keywords and phrases.} \small Nonlinear
differential equation, nonlinear partial differential equation,
seventh-order evolution equation, KdV7, soliton solution,
Kaup-Kupershmidt equation.\par}
\smallskip {\narrower \noindent {\itp 2000 Mathematics Subject
 Classification.} \small 35C05. \par}
\normalsize

\vspace{0.5cm}

\section{Introduction}
The general form of the seventh order KdV (KdV7) \cite{hereman}
reads
\begin{equation}\label{eq01}
u_t + a u^3 u_x + b u_x^3+c u u_x u_{xx}+d u^2 u_{xxx}+e
u_{2x}u_{3x}+f u_x u_{4x}+g u u_{5x}+u_{7x}=0.
\end{equation}
The seventh-order KdV equation has been introduced by Pomeau et al.
\cite{pomeau} for discussing the structural stability of KdV
equation under a singular perturbation. Some particular cases of Eq.
(\ref{eq01}) are :
\begin{itemize}
\item Seventh-order Sawada-Kotera-Ito equation \cite{hereman}
( $a=252$, $b=63$, $c=378$, $d=126$, $e=63$, $f=42$, $g=21$ ) :
\begin{equation}\label{eq02}
u_t + 252 u^3 u_x + 63 u_x^3+378 u u_x u_{xx}+126 u^2 u_{xxx}+63
u_{2x}u_{3x}+42 u_x u_{4x}+21 u u_{5x}+u_{7x}=0.
\end{equation}

\item Seventh-order Lax equation \cite{hereman}
( $a=140$, $b=70$, $c=280$, $d=70$, $e=70$, $f=42$, $g=14$ ) :
\begin{equation}\label{eq03}
u_t + 140 u^3 u_x + 70 u_x^3+280 u u_x u_{xx}+70 u^2 u_{xxx}+70
u_{2x}u_{3x}+42 u_x u_{4x}+14 u u_{5x}+u_{7x}=0.
\end{equation}

\item Seventh-order Kaup-Kupershmidt equation \cite{hereman}
( $a=2016$, $b=630$, $c=2268$, $d=504$, $e=252$, $f=147$, $g=42$ ) :
\begin{equation}\label{eq03a}
u_t + 2016 u^3 u_x + 630 u_x^3+2268 u u_x u_{xx}+504 u^2 u_{xxx}+252
u_{2x}u_{3x}+147 u_x u_{4x}+42 u u_{5x}+u_{7x}=0.
\end{equation}
\end{itemize}
In the next sections we obtain some exact solutions for the
seventh-order Kaup-Kupershmidt equation (\ref{eq03a}).
\section{Soliton solutions by the Cole-Hopf transformation}
Some nonlinear pse's in the variable $u=u(x,t)$ may be solved via
the substitution
\begin{equation}\label{cole}
u(x,t)=A\partial_{xx}\log(1+f(x,t))+B,
\end{equation}
where $A$ and $B$ are some constants and $f(x,t)$ is a new unknown
function. Eq. (\ref{cole}) is called a generalized Cole-Hopf
transformation \cite{cole}\cite{hopf}. We will solve Eq.
(\ref{eq03a}) by using (\ref{cole}) for the special choice
$$f(x,t)=\exp(k x-c t+\delta),$$
being $c$, $k$ and $\delta$ some constants.  Now, we make the
substitution
\begin{equation}\label{cole1}
u(x,t)=A\partial_{xx}\log(1+\exp(k x-c t+\delta))+B
\end{equation}
into (\ref{eq03a}). We obtain a polynomial equation in the variable
$\zeta=\exp(kx- c t+\delta)$. Equating the coefficients of the
different powers of $\zeta$ to zero, we obtain the following
algebraic system :
\begin{itemize}
\item[$\bullet$] $-A k^9-42 A B k^7-504 A
   B^2 k^5-2016 A B^3 k^3+A c
   k^2=0.$  \item[$\bullet$] $A k^9+42 A B k^7+504
   A B^2 k^5+2016 A B^3 k^3-A
   c k^2=0.$  \item[$\bullet$] $-441 A^2 k^9+247 A
   k^9-3276 A^2 B k^7+2310 A B
   k^7-6048 A^2 B^2 k^5+3528 A
   B^2 k^5-10080 A B^3 k^3+5 A
   c k^2=0.$  \item[$\bullet$] $441 A^2 k^9-247 A
   k^9+3276 A^2 B k^7-2310 A B
   k^7+6048 A^2 B^2 k^5-3528 A
   B^2 k^5+10080 A B^3 k^3-5 A
   c k^2=0.$  \item[$\bullet$] $-3402 A^3 k^9+10143
   A^2 k^9-4293 A k^9-6048 A^3
   B k^7+15876 A^2 B k^7-7938
   A B k^7-18144 A^2 B^2
   k^5+13608 A B^2 k^5-18144 A
   B^3 k^3+9 A c k^2=0.$  \item[$\bullet$] $3402
   A^3 k^9-10143 A^2 k^9+4293
   A k^9+6048 A^3 B k^7-15876
   A^2 B k^7+7938 A B
   k^7+18144 A^2 B^2 k^5-13608
   A B^2 k^5+18144 A B^3 k^3-9
   A c k^2=0.$  \item[$\bullet$] $-2016 A^4
   k^9+18774 A^3 k^9-40320 A^2
   k^9+15619 A k^9-6048 A^3 B
   k^7+19152 A^2 B k^7-10290 A
   B k^7-12096 A^2 B^2
   k^5+9576 A B^2 k^5-10080 A
   B^3 k^3+5 A c k^2=0.$  \item[$\bullet$] $2016
   A^4 k^9-18774 A^3 k^9+40320
   A^2 k^9-15619 A k^9+6048
   A^3 B k^7-19152 A^2 B
   k^7+10290 A B k^7+12096 A^2
   B^2 k^5-9576 A B^2
   k^5+10080 A B^3 k^3-5 A c
   k^2=0.$
   \end{itemize}
Solving this systems gives
\begin{equation}\label{solcole}
A=\dfrac{1}{2},\qquad B=-\dfrac{k^2}{24},\qquad c=-\dfrac{k^7}{48}.
\end{equation}
Therefore, a one-soliton solution of (\ref{eq03a}) is given by
\begin{equation}\label{onesoli}
u_0(x,t)=-\frac{\left(1-10 e^{\frac{t
   k^7}{48}+x k+\delta
   }+e^{\frac{t k^7}{24}+2 x
   k+2 \delta }\right) k^2}{24
   \left(1+e^{\frac{t
   k^7}{48}+x k+\delta
   }\right)^2},
\end{equation}
which simplifies to
\begin{equation}
u_{0}(x,t)=-\frac{k^{2}}{24}+\frac{1}{4(1+\cosh \left(
kx+\frac{k^{7}}{48} t+\delta \right) )}.  \label{onesoli1}
\end{equation}

\section{Exact solutions via two rational hyperbolic ansatze}
Using the wave transformation
\begin{equation}
u(x,t)=v(\xi),\,\,\xi=x+\lambda t, \label{eq04}
\end{equation}
where $\lambda$ is a constant, Eq. (\ref{eq03a}) becomes the
nonlinear ode
\begin{equation}
\begin{tabular}{ll}
$2016 v'(\xi ) u(\xi )^3+504
   v^{(3)}(\xi ) v(\xi
   )^2+2268 v'(\xi ) v''(\xi )
   v(\xi )+42 v^{(5)}(\xi )
   v(\xi )+630 v'(\xi
   )^3+$
&  \\
$\lambda  v'(\xi )+252
   v''(\xi ) v^{(3)}(\xi )+147
   v'(\xi ) v^{(4)}(\xi
   )+v^{(7)}(\xi ) =0.$ &
\end{tabular}
\label{eq06}
\end{equation}
We shall find solutions of Eq. (\ref{eq06}) by using two ansatze:
\begin{itemize}
\item \textbf{The tanh--coth ansatz}:
\begin{equation}\label{eq07}
v(\xi)=p+a\tanh(\mu \xi)+b\coth(\mu\xi)+c\tanh^2(\mu
\xi)+d\coth^2(\mu\xi)
\end{equation}
\item \textbf{A sinh--cosh rational ansatz}:
\begin{equation}\label{eq08}
v(\xi)=p+\dfrac{k}{1+c\sinh(\mu\xi)+d \cosh(\mu\xi)},
\end{equation}
\end{itemize}
where $a$, $b$, $c$, $d$, $k$, $p$ and $\mu$ are constants. Some
times, by replacing $\mu$ by $\sqrt{-1}\mu$, we obtain periodic
solutions. \subsection{Solutions by the tanh--coth ansatz} We change
the $\tanh$ and $\coth$ functions to their exponential form and then
we substitute (\ref{eq07}) into (\ref{eq06}). We obtain a polynomial
equation in the variable $\zeta=\exp(\mu\xi)$. Equating the
coefficients of the different powers of $\zeta$ to zero results in
an algebraic system in the variables $a$, $b$, $c$, $d$,
 $p$, $\lambda$ and $\mu$. Solving it with the aid of a
computer, we obtain following solutions of (\ref{eq03a}) :

\begin{itemize}
\item[$\bullet $] $a=0,$ $\,\,$ $b=0,$ $\,\,$ $c=0,$ $\,\,$ $d=-\frac{\mu
^{2}}{2},$ $\,\,$ $p=\frac{\mu ^{2}}{3},$ $\,\,$ $\lambda
=\frac{4\mu ^{6}}{3 }:$
\begin{equation*}
u_{1}(x,t)=\frac{\mu ^{2}}{3}-\frac{\mu ^{2}}{2}\coth ^{2}\left( \mu
\left( x+\frac{4\mu ^{6}}{3}t\right) \right) .
\end{equation*}

\begin{equation*}
u_{2}(x,t)=-\frac{\mu ^{2}}{3}-\frac{\mu ^{2}}{2}\cot ^{2}\left( \mu
\left( x-\frac{4\mu ^{6}}{3}t\right) \right) .
\end{equation*}

\item[$\bullet $] $a=0,$ $\,\,$ $b=0,$ $\,\,$ $c=-\frac{\mu ^{2}}{2},$ $\,\,$
$d=0,$ $\,\,$ $p=\frac{\mu ^{2}}{3},$ $\,\,$ $\lambda =\frac{4\mu
^{6}}{3}:$
\begin{equation*}
u_{3}(x,t)=\frac{\mu ^{2}}{3}-\frac{\mu ^{2}}{2}\tanh ^{2}\left( \mu
\left( x+\frac{4\mu ^{6}}{3}t\right) \right) .
\end{equation*}
\begin{equation*}
u_{4}(x,t)=-\frac{\mu ^{2}}{3}-\frac{\mu ^{2}}{2}\tan ^{2}\left( \mu
\left( x-\frac{4\mu ^{6}}{3}t\right) \right) .
\end{equation*}

\item[$\bullet $] $a=0,$ $\,\,$ $b=0,$ $\,\,$ $c=-\frac{\mu ^{2}}{2},$ $\,\,$
$d=-\frac{\mu ^{2}}{2},$ $\,\,$ $p=\frac{\mu ^{2}}{3},$ $\,\,$
$\lambda = \frac{256\mu ^{6}}{3}:$

\begin{equation*}
u_{5}(x,t)=\frac{\mu ^{2}}{3}-\frac{\mu ^{2}}{2}\coth ^{2}\left( \mu
\left( x+\frac{256\mu ^{6}}{3}t\right) \right) -\frac{\mu
^{2}}{2}\tanh ^{2}\left( \mu \left( x+\frac{256\mu ^{6}}{3}t\right)
\right) .
\end{equation*}
\begin{equation*}
u_{6}(x,t)=-\frac{\mu ^{2}}{3}-\frac{\mu ^{2}}{2}\cot ^{2}\left( \mu
\left( x-\frac{256\mu ^{6}}{3}t\right) \right) -\frac{\mu
^{2}}{2}\tan ^{2}\left( \mu \left( x-\frac{256\mu ^{6}}{3}t\right)
\right) .
\end{equation*}
\end{itemize}
\subsection{Solutions by a sinh--cosh rational ansatz} We change the
$\sinh$ and $\cosh$ functions to their exponential form and then we
substitute (\ref{eq08}) into (\ref{eq06}). As a result, we obtain a
polynomial equation in the variable $\zeta=\exp(\mu\xi)$. Equating
the coefficients of the different powers of $\zeta$ to zero results
in an algebraic system in the variables $c$, $d$,
 $k$, $p$, $\lambda$ and $\mu$. Solving it with the aid of a
computer yields following solutions of (\ref{eq03a}) :
\begin{itemize}
\item[$\bullet $] $c=0,$ $\,\,$ $d=-1,$ $\,\,$ $k=\frac{\mu ^{2}}{4},$ $\,\,$
$p=-\frac{\mu ^{2}}{24},$ $\,\,$ $\lambda =\frac{\mu ^{6}}{48}:$
\begin{equation*}
u_{7}(x,t)=-\frac{\mu ^{2}}{24}+\frac{\mu ^{2}}{4\left( 1-\cosh
\left( \mu \left( x+\frac{\mu ^{6}}{48}t\right) \right) \right) }.
\end{equation*}
\begin{equation*}
u_{8}(x,t)=\frac{\mu ^{2}}{24}-\frac{\mu ^{2}}{4\left( 1-\cos \left(
\mu \left( x-\frac{\mu ^{6}}{48}t\right) \right) \right) }.
\end{equation*}

\item[$\bullet $] $c=0,$ $\,\,$ $d=1,$ $\,\,$ $k=\frac{\mu ^{2}}{4},$ $\,\,$
$p=-\frac{\mu ^{2}}{24},$ $\,\,$ $\lambda =\frac{\mu ^{6}}{48}:$
\begin{equation*}
u_{9}(x,t)=-\frac{\mu ^{2}}{24}+\frac{\mu ^{2}}{4\left( 1+\cosh
\left( \mu \left( x+\frac{\mu ^{6}}{48}t\right) \right) \right) }.
\end{equation*}
\begin{equation*}
u_{10}(x,t)=\frac{\mu ^{2}}{24}-\frac{\mu ^{2}}{4\left( 1+\cos
\left( \mu \left( x-\frac{\mu ^{6}}{48}t\right) \right) \right) }.
\end{equation*}

\item[$\bullet $] $c=-\sqrt{-1},$ $\,\,$ $d=0,$ $\,\,$ $k=\frac{\mu ^{2}}{4},
$ $\,\,$ $p=-\frac{\mu ^{2}}{24},$ $\,\,$ $\lambda =\frac{\mu
^{6}}{48}:$
\begin{equation*}
u_{11}(x,t)=-\frac{\mu ^{2}}{24}+\frac{\mu ^{2}}{4\left(
1-\sqrt{-1}\sinh \left( \mu \left( x+\frac{\mu ^{6}}{48}t\right)
\right) \right) }.
\end{equation*}
\begin{equation*}
u_{12}(x,t)=\frac{\mu ^{2}}{24}-\frac{\mu ^{2}}{4\left( 1+\sin
\left( \mu \left( x-\frac{\mu ^{6}}{48}t\right) \right) \right) }.
\end{equation*}

\item[$\bullet $] $c=\sqrt{-1},$ $\,\,$ $d=0,$ $\,\,$ $k=\frac{\mu ^{2}}{4},$
$\,\,$ $p=-\frac{\mu ^{2}}{24},$ $\,\,$ $\lambda =\frac{\mu
^{6}}{48}:$
\begin{equation*}
u_{13}(x,t)=-\frac{\mu ^{2}}{24}+\frac{\mu ^{2}}{4\left(
1+\sqrt{-1}\sinh \left( \mu \left( x+\frac{\mu ^{6}}{48}t\right)
\right) \right) }.
\end{equation*}
\begin{equation*}
u_{14}(x,t)=\frac{\mu ^{2}}{24}-\frac{\mu ^{2}}{4\left( 1-\sin
\left( \mu \left( x-\frac{\mu ^{6}}{48}t\right) \right) \right) }.
\end{equation*}

\item[$\bullet $] $c=\sqrt{d^{2}-1},$ $\,\,$ $k=\frac{\ mu^{2}}{4},$ $\,\,$ $
p=-\frac{\mu ^{2}}{24},$ $\,\,$ $\lambda =\frac{\mu ^{6}}{48}:$
\begin{equation*}
u_{15}(x,t)=-\frac{\mu ^{2}}{24}+\frac{\mu ^{2}}{4\left( 1+d\cosh
\left( \mu \left( x+\frac{\mu ^{6}}{48}t\right) \right)
+\sqrt{d^{2}-1}\sinh \left( \mu \left( x+\frac{\mu ^{6}}{48}t\right)
\right) \right) }.
\end{equation*}
\begin{equation*}
u_{16}(x,t)=\frac{\mu ^{2}}{24}-\frac{\mu ^{2}}{4\left( 1+d\cos
\left( \mu \left( x-\frac{\mu ^{6}}{48}t\right) \right)
+\sqrt{1-d^{2}}\sin \left( \mu \left( x-\frac{\mu ^{6}}{48}t\right)
\right) \right) }.
\end{equation*}

\item[$\bullet $] $c=-\sqrt{d^{2}-1},$ $\,\,$ $k=\frac{\mu ^{2}}{4},$ $\,\,$
$p=-\frac{\mu ^{2}}{24},$ $\,\,$ $\lambda =\frac{\mu ^{6}}{48}:$
\begin{equation*}
u_{17}(x,t)=-\frac{\mu ^{2}}{24}+\frac{\mu ^{2}}{4\left( 1+d\cosh
\left( \mu \left( x+\frac{\mu ^{6}}{48}t\right) \right)
-\sqrt{d^{2}-1}\sinh \left( \mu \left( x+\frac{\mu ^{6}}{48}t\right)
\right) \right) }.
\end{equation*}
\begin{equation*}
u_{18}(x,t)=\frac{\mu ^{2}}{24}-\frac{\mu ^{2}}{4\left( 1+d\cos
\left( \mu \left( x-\frac{\mu ^{6}}{48}t\right) \right)
-\sqrt{1-d^{2}}\sin \left( \mu \left( x-\frac{\mu ^{6}}{48}t\right)
\right) \right) }.
\end{equation*}
\end{itemize}
\section{Conclusions} We obtained solutions for the seventh-order KK
equation (\ref{eq03a}) by using three distinct methods. These
methods are direct and effective and they are applicable to solve
other seventh-order KdV equations. We think that the results given
here are new in the literature.

\end{document}